\documentclass[rnote]{aa} % for the research notes
\usepackage{graphicx}
\usepackage{txfonts}
\usepackage{natbib}
\bibpunct{(}{)}{;}{a}{}{,} % to follow the A&A style
\usepackage{hyperref}
\usepackage{url}

\newcommand{\te}{\ensuremath{T_{\mathrm{eff}}}}
\newcommand{\teo}{\ensuremath{T_{\mathrm{o}}}}

\newcommand{\kep}{\ensuremath{_{\mathrm{K}}}}
\newcommand{\bol}{\ensuremath{_{\mathrm{b}}}}
\newcommand{\dmu}{\ensuremath{\,\mathrm{d}\mu}}
\newcommand{\dlam}{\ensuremath{\,\mathrm{d}\lambda}}
\newcommand{\Vtot}{\ensuremath{\tilde V_{\mathrm{tot}}}}
\newcommand{\Vl}{\ensuremath{\tilde V_\ell}}
\newcommand{\TK}{\ensuremath{{\cal T}\kep}}

\begin{document}
\title{Visibilities and bolometric corrections for stellar oscillation modes observed by Kepler\thanks{Full Table 1 is only available in electronic form at the CDS via anonymous ftp to \url{cdsarc.u-strasbg.fr} (\url{130.79.128.5}) or via \url{http://cdsweb.u-strasbg.fr/cgi-bin/qcat?J/A+A/531/A124}}}

   \author{%
J. Ballot\inst{\ref{inst:irap1},\ref{inst:irap2}} \and
C. Barban\inst{\ref{inst:lesia}} \and
C. van 't Veer-Menneret\inst{\ref{inst:gepi}}
}
   \institute {%
CNRS, Institut de Recherche en Astrophysique et Plan\'etologie, 14 avenue Edouard Belin, 31400 Toulouse, France\label{inst:irap1}\\
\email{jballot@ast.obs-mip.fr}
\and
Universit\'e de Toulouse, UPS-OMP, IRAP, Toulouse, France\label{inst:irap2}
\and
Observatoire de Paris, LESIA, CNRS UMR 8109, Universit\'e Pierre et Marie Curie, Universit\'e Denis Diderot, 5 Place J. Janssen, 92195 Meudon, France\label{inst:lesia}
\and 
Observatoire de Paris, GEPI, CNRS UMR 8111, Universit\'e Denis Diderot,  5 Place J. Janssen, 92195 Meudon Principal Cedex, France\label{inst:gepi}}

   \date{Received 29 November 2010 / Accepted 23 May 2011}

 \abstract{Kepler produces a large amount of data used for asteroseismological analyses, particularly of solar-like stars and red giants. The mode amplitudes observed in the Kepler spectral band have to be converted into bolometric amplitudes to be compared to models.}%
{We give a simple bolometric correction for the amplitudes of radial modes observed with Kepler, as well as the relative visibilities of non-radial modes.}%
{We numerically compute the bolometric correction $c_{\mathrm{K-bol}}$ and mode visibilities for different effective temperatures $\te$ within the range 4000--7500~K, using a similar approach to a recent one from the literature.}%
{We derive a law for the correction to bolometric values: $c_{\mathrm{K-bol}}=1+a_1(\te-\teo)+a_2(\te-\teo)^2$, with $\teo=5934$~K, $a_1=1.349\times10^{-4}$~K${}^{-1}$, and $a_2=-3.120\times10^{-9}$~K${}^{-2}$ or, alternatively, as the power law $c_{\mathrm{K-bol}}=(\te/\teo)^\alpha$ with $\alpha=0.80$. We give tabulated values for the mode visibilities based on limb-darkening (LD), computed from ATLAS9 model atmospheres for $\te \in [4000, 7500]$~K, $\log g \in [2.5, 4.5]$, and [M/H]${} \in [-1.0,+1.0]$. We show that using LD profiles already integrated over the spectral band provides quick and good approximations for visibilities. We point out the limits of these classical visibility estimations.}%
{} 
% 5 {} token are mandatory
 

\maketitle
%
%________________________________________________________________

\section{Introduction}
While the NASA Kepler mission \citep[e.g.][]{Koch10} is dedicated to exoplanet finding, the high-quality photometric data provided by the instrument are perfectly suited for asteroseismology, specially for the study of solar-like oscillations in main-sequence and red giant stars \citep[e.g.][]{Chaplin10,Bedding10}.
For solar-like oscillations, the characteristics of acoustic modes, such as their frequencies, amplitudes, or lifetimes are determined by fitting the oscillation power spectrum \citep[see, e.g.,][and references therein for a description of these techniques]{Ballot10}. The apparent amplitudes of modes recovered in the power spectrum -- or the light curve -- depend on the spectral response of the instrument. It is therefore fundamental to recover the bolometric amplitudes of modes to be able, for example, to compare the observed amplitudes to theoretical predictions.

Moreover, owing to cancellation effects, the apparent amplitude of a mode depends on its degree $\ell$. Modes with degrees $\ell>3$ are generally considered to be invisible. The knowledge of these mode visibilities is often requested for solar-like oscillation analyses. They sensitively depend on the stellar limb darkening (LD), which depends on the observed wavelengths and requests a knowledge of stellar atmospheres.

In this note, we first propose in Sect.~\ref{sec:cbol} a simple bolometric correction for radial modes observed with Kepler, following an approach similar to the one developed for the Convection, Rotation, and planetary Transits \citep[CoRoT,][]{Baglin06} satellite by \citet[][hereafter M09]{Michel09}. In Sect.~\ref{sec:vis} we give the visibilities of non-radial modes for Kepler, using LD profiles computed from ATLAS9 model atmospheres in a modified version for the convection treatment, and we discuss the use of LD laws integrated over spectral bands. Finally, we discuss in Sect.~\ref{sec:dis} the use of these correcting factors, and point out some of their limits.

\section{Bolometric correction for radial modes}\label{sec:cbol}

We derive in this part the bolometric correction for radial modes. This correction is the factor to apply to the amplitude of a radial mode observed in a given spectral band to recover the bolometric amplitude.

\begin{figure}[t]
    \centering
    \includegraphics[width=\hsize]{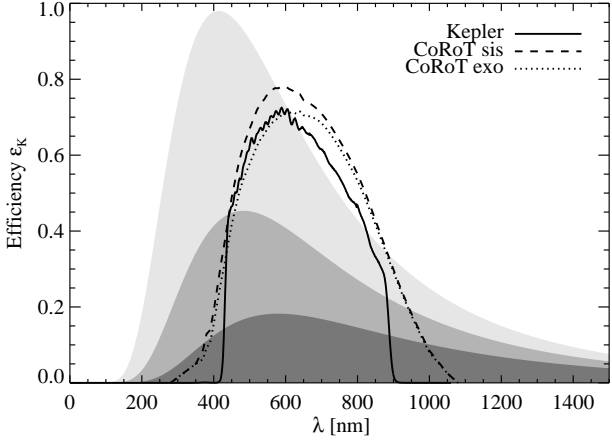}
    \caption{A solid line shows the spectral response ${\cal E}\kep$ of Kepler as a function of the wavelength $\lambda$ \citep[after][]{KIH}, whereas dotted and dashed lines show the spectral responses of CoRoT seismo- and exofield \citep{Auvergne09}. Grey profiles indicate the spectra (in arbitrary units) of black bodies with temperatures of 7000, 6000, and 5000~K, from top to bottom.
    \label{Fig:Response}}
\end{figure}

The fluctuations induced by \emph{radial} modes identically affect the whole atmosphere, and can be considered as fluctuations $\delta T$ of the effective temperature $\te$.
For photometric measurements, made for example with Kepler, the oscillations generate fluctuations $\delta F\kep$ of the measured total flux $F\kep$. This flux depends on the response of the instrument.
By considering fluctuations small enough to be considered as linear perturbations, and by assuming that the observed star radiates as a black body of temperature $\te$, the fluctuations $\delta F\kep$ and $\delta T$ are linked through the relation:
\begin{equation}
\frac{\delta F\kep}{F\kep} = \frac%
{\int \TK(\lambda) \frac{\partial B}{\partial \te}(\lambda,\te) \delta T \dlam}%
{\int \TK(\lambda) B(\lambda,\te) \dlam},\label{eq:dFk_Fk}
\end{equation}
where $B(\lambda,\te)$ is the Planck function, and $\TK(\lambda)$ is the transfer function of the instrument, and $\lambda$ the light wavelength (see M09 for a detailed derivation).
The transfer function reads $\TK(\lambda)={\cal E}\kep(\lambda)/E_\nu$, with ${\cal E}\kep$ the spectral response of the detector, and $E_\nu=hc/\lambda$ the photon energy at the wavelength $\lambda$. The constants $h$ and $c$ denote the Planck constant and the light speed. The spectral response of Kepler is described in \citet{KIH}\footnote{\url{http://keplergo.arc.nasa.gov/CalibrationResponse.shtml}} and plotted in Fig.~\ref{Fig:Response}. For comparison purposes, the responses of the two CoRoT channels \citep[see][]{Auvergne09} are also plotted.

The fluctuation $\delta F\bol$ of the bolometric flux $F\bol$ is simply linked with $\delta T$:
\begin{equation}
\frac{\delta F\bol}{F\bol} = 4\frac{\delta T}{\te}.
\end{equation}
We then deduce
\begin{equation}
\frac{\delta F\bol}{F\bol} = c_{\mathrm{K-bol}}(\te)
\frac{\delta F\kep}{F\kep}
\end{equation}
with the bolometric correction factor
\begin{equation}
c_{\mathrm{K-bol}}(\te) =
\frac{4 \int \TK(\lambda) B(\lambda,\te) \dlam}%
{\te \int \TK(\lambda) \frac{\partial B}{\partial \te}(\lambda,\te)\dlam}. \label{eq:ckbol}
\end{equation}

\begin{figure}[t]
    \centering
    \includegraphics[width=\hsize]{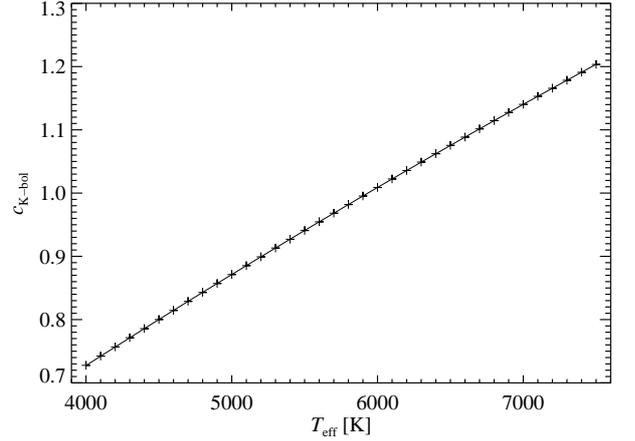}
    \caption{Bolometric correction $c_{\mathrm{K-bol}}$ as a function of the effective temperature $\te$. Crosses are numerical calculations of Eq.~\ref{eq:ckbol}, the line is a second-order polynomial fit of these points.
      \label{Fig:ckbol}}
\end{figure}

We numerically computed the bolometric corrections for effective temperatures $\te$ within the range 4000--7500~K, covering the range of solar-like oscillating stars.
Results are shown in Fig.~\ref{Fig:ckbol}. A second-order polynomial fairly fits the computed set of points. We then obtain the following law:
\begin{equation}
c_{\mathrm{K-bol}}(\te)=1+a_1(\te-\teo)+a_2(\te-\teo)^2
\end{equation}
 with the coefficients $\teo=5934$~K, $a_1=1.349\times10^{-4}$~K${}^{-1}$, and $a_2=-3.12\times10^{-9}$~K${}^{-2}$.
This approximation, also plotted in Fig.~\ref{Fig:ckbol}, gives consistent results with the numerical computations within $\sim 10^{-4}$. This error is probably negligible compared to the method error mainly induced by the departure of stellar spectra from black body emissions.

Alternatively, we express the bolometric correction as a power law
\begin{equation}
c_{\mathrm{K-bol}}(\te)=(\te/\teo)^\alpha
\end{equation}
with $\alpha=0.80$. This approximation is consistent with the numerical computations within $\sim 10^{-3}$ over the considered temperature range.

\section{Visibilities of non-radial modes}\label{sec:vis}

\begin{figure*}[!ht]
    \centering
    \includegraphics[width=\hsize]{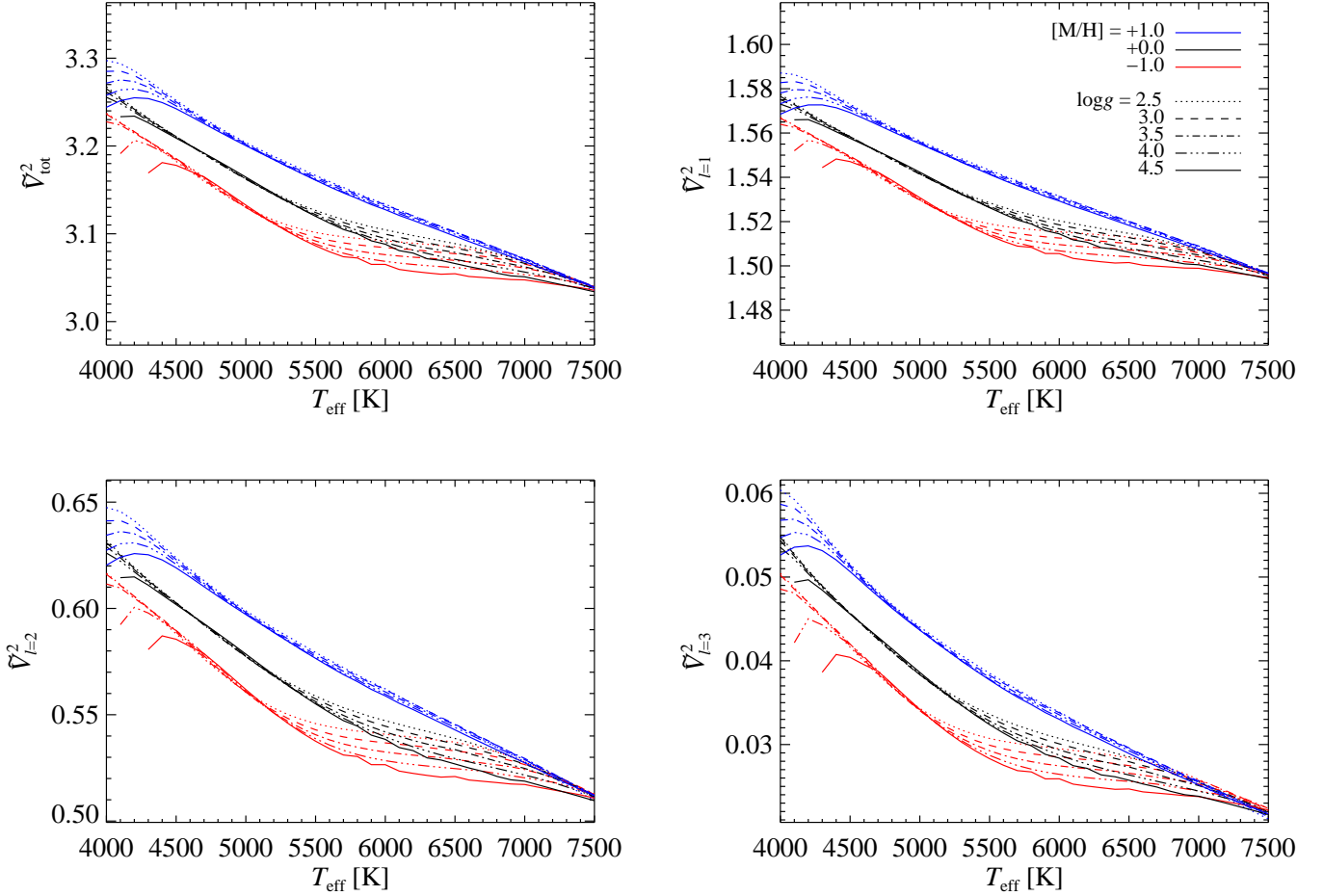}
    \caption{Total visibility $\Vtot^2$ and relative visibilities $\Vl^2$ for degrees $\ell=1,2$, and 3 plotted as a function of the effective temperature $\te$, for different surface gravities $\log g$ (plotted with different line styles), and different metallicities [M/H] (plotted with different colours).
      \label{Fig:vis}}
\end{figure*}

Computing the relative visibilities of non-radial modes compared to $\ell=0$ modes requires the knowledge of the stellar limb darkening. For a monochromatic observation at the wavelength $\lambda$, the visibility $V_\ell$ of a mode of degree $\ell$ reads \citep[e.g.][]{Dziembowski77,Gizon03}
\begin{equation}
  \label{eq:Vl}
  V_\ell= \sqrt{\pi(2\ell+1)} \int_0^1 P_\ell(\mu) W(\mu) \mu \dmu
\end{equation}
where $P_\ell$ is the $\ell$th-order Legendre polynomial and $W(\mu)$ is a weighting function depending on $\mu$, the distance to the limb (equal to 0 at the limb and 1 at the disc centre), defined as the cosine of the angle $\widehat{MCO}$,  where $M$ is the considered point at the stellar surface, $C$ the centre of the star, and $O$ the observer.
The function $W(\mu)$ is identified to the LD profile $g_\lambda(\mu)=I_\lambda(\mu)/I_\lambda(1)$, where $I_\lambda(\mu)$ is the specific intensity at the wavelength $\lambda$.

We also define the normalised --or relative-- visibilities $\Vl=V_\ell/V_0$, and the total visibility
\begin{equation}
\Vtot^2=\sum_{\ell=0}^\infty \Vl^2.
\end{equation}
We show \citep[after][Eq. A.16]{Ballot10} that
\begin{equation}
\sum_{\ell=0}^\infty V_\ell^2=
2\pi \int_0^1W(\mu)^2\mu^2 \dmu.
\end{equation}
Thus, we deduce the total visibility
\begin{equation}
\Vtot^2=\frac%
{2 \int_0^1W(\mu)^2\mu^2 \dmu}%
{\left(\int_0^1W(\mu)\mu \dmu\right)^2}.\label{eq:Vtot}
\end{equation}

For a broad band observation, the computation is not as simple. In M09 an expression for the mode visibilities is derived by following the approach of \citet{Berthomieu90}. It is straightforward to reformulate their expression (M09, Eq. A.18) in the same form as Eq.~\ref{eq:Vl} by considering $W(\mu)=W\kep (\mu)$
with
\begin{equation}
W\kep (\mu) = \frac%
{\int 
 \TK(\lambda)
 \te\frac{\partial B}{\partial \te}(\lambda,\te)
 H_\lambda g_\lambda(\mu) \dlam}%
{\int \TK(\lambda) B(\lambda,\te) H_\lambda G_\lambda \dlam},
\end{equation}
where
\begin{equation}
 G_\lambda = \int_0^1 \mu g_\lambda(\mu) \dmu
\quad\mbox{and}\quad
 H_\lambda = \left(\int_0^1 g_\lambda(\mu) \dmu\right)^{-1}.
\end{equation}

\begin{table}[ht]
  \centering
  \caption{Total visibility $\Vtot^2$ and relative visibilities $\Vl^2$ for degrees $\ell=1,2$, and 3 for different fundamental stellar parameters $\te$, $\log g$, and [M/H].\label{tab:kepler}}

  \begin{tabular}{ccccccc}
\hline\hline
$\te$ [K] & $\log g$ & [M/H] & $\Vtot^2$ &
$\tilde V_1^2$ & $\tilde V_2^2$ & $\tilde V_3^2$ \\
\hline
 4600 & 4.5 & $-1.0$ & 3.17 & 1.54 & 0.58 & 0.040 \\
 4600 & 4.5 & $+0.0$ & 3.20 & 1.55 & 0.60 & 0.044 \\
 4600 & 4.5 & $+1.0$ & 3.23 & 1.57 & 0.61 & 0.049 \\
 5000 & 4.5 & $-1.0$ & 3.13 & 1.53 & 0.56 & 0.034 \\
 5000 & 4.5 & $+0.0$ & 3.16 & 1.54 & 0.58 & 0.039 \\
 5000 & 4.5 & $+1.0$ & 3.20 & 1.56 & 0.60 & 0.044 \\
 5400 & 4.5 & $-1.0$ & 3.09 & 1.52 & 0.54 & 0.029 \\
 5400 & 4.5 & $+0.0$ & 3.13 & 1.53 & 0.56 & 0.033 \\
 5400 & 4.5 & $+1.0$ & 3.17 & 1.54 & 0.58 & 0.039 \\
 5800 & 4.5 & $-1.0$ & 3.07 & 1.51 & 0.53 & 0.027 \\
 5800 & 4.5 & $+0.0$ & 3.10 & 1.52 & 0.54 & 0.030 \\
 5800 & 4.5 & $+1.0$ & 3.14 & 1.53 & 0.57 & 0.035 \\
 6200 & 4.5 & $-1.0$ & 3.06 & 1.50 & 0.52 & 0.025 \\
 6200 & 4.5 & $+0.0$ & 3.08 & 1.51 & 0.53 & 0.027 \\
 6200 & 4.5 & $+1.0$ & 3.12 & 1.52 & 0.55 & 0.031 \\
 6600 & 4.5 & $-1.0$ & 3.05 & 1.50 & 0.52 & 0.024 \\
 6600 & 4.5 & $+0.0$ & 3.06 & 1.50 & 0.52 & 0.025 \\
 6600 & 4.5 & $+1.0$ & 3.09 & 1.52 & 0.54 & 0.028 \\
 7000 & 4.5 & $-1.0$ & 3.05 & 1.50 & 0.52 & 0.024 \\
 7000 & 4.5 & $+0.0$ & 3.05 & 1.50 & 0.52 & 0.024 \\
 7000 & 4.5 & $+1.0$ & 3.07 & 1.51 & 0.53 & 0.025 \\
\hline
  \end{tabular}
\tablefoot{The full table can be found at the CDS, or at \url{http://www.ast.obs-mip.fr/article998.html}, or at  \url{http://www.jerome-ballot.fr/}.}
\end{table}

We computed LD profiles $g_\lambda(\mu)$ as described in \citet{Barban03}. For this, we used stellar atmosphere models computed with the ATLAS9 code\footnote{\url{http://kurucz.harvard.edu}} \citep{Kurucz93} in a modified version including the convective prescription of \citet{Canuto96}, known as CGM \citep[for details, see][]{Heiter02}.
We then computed $\Vtot^2$, as well as $\Vl^2$ for $\ell=1,2$ and 3, for a large range of stellar fundamental parameters: for effective temperatures $\te \in [4000, 7500]$~K, surface gravities $\log g \in [2.5, 4.5]$, and three metallicities\footnote{We recall that [X/Y] means $\log(X/Y)_\mathrm{star}-\log(X/Y)_{\sun}$, $X$ and $Y$ being chemical element abundances in number of atoms per unit volume.} [M/H]${}=-1.0,+0.0,$ and $+1.0$. Results are listed in Table~\ref{tab:kepler}, and are plotted in Fig.~\ref{Fig:vis}. As expected, it mainly depends on the temperature $\te$, while the sensitivity to $\log g$ is weak. However, strong changes in metallicity have visible impacts on the results.

\begin{figure}[ht]
    \centering
     \includegraphics[width=\hsize]{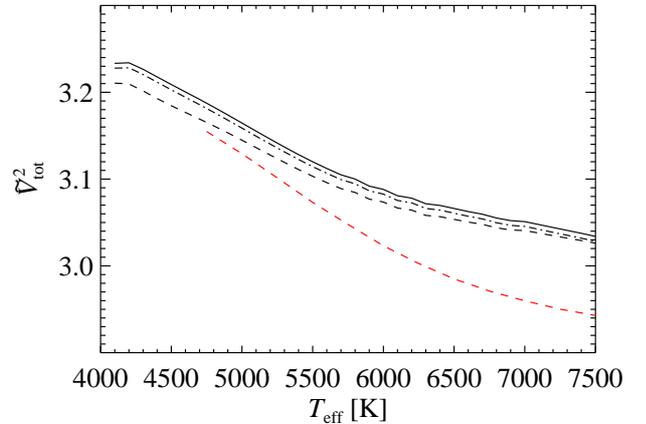}
    \caption{Total visibility $\Vtot^2$ plotted as a function of the effective temperature $\te$, for $\log g=4.5$ and [M/H]${}=+0.0$. The solid line shows the complete computation, the dotted-dashed line corresponds to the simplification of M09, the dashed line shows the approximated results obtained by using integrated $g\kep$ LD profiles. The red dashed line is obtained by using the LD laws published by \citet{Sing10}.
      \label{Fig:comp}}
\end{figure}

In M09, a simplification has been done by noticing\footnote{This simplification was also made to obtain Eq.~\ref{eq:dFk_Fk}, see M09.} that the product $H_\lambda G_\lambda$ varies slowly with $\lambda$. Moreover, $\Vtot^2$ is not obtained through Eq.~\ref{eq:Vtot}, but computed as the truncated sum $\sum_{\ell=0}^4 \Vl^2$. As shown in Fig.~\ref{Fig:comp} for $\log g = 4.5$ and [M/H]${}=+0.0$, the results of this simplification give very close results to the more complete computations.

Even by considering this simplification, the calculations of \Vl\ and \Vtot\ require one to know wavelength-dependent LD laws $g_\lambda (\mu)$, of which computations are time-consuming. However, we find in the literature LD profiles which are already integrated over spectral bands of interest \citep[for Kepler, see for example][]{Sing10}. We denote $g\kep (\mu) = I\kep (\mu)/I\kep (1)$ the integrated LD profile where
\begin{equation}
 I\kep (\mu) =\int \TK(\lambda) I_\lambda(\mu) \dlam.
\end{equation}
We propose to use Eq.~\ref{eq:Vl} with $W(\mu)=g\kep(\mu)$ to perform approximated computations of \Vl\ and \Vtot. Formally, this approximation is a correct simplification of the full computation for narrow-band filters only. Because the Kepler band is fairly broad, this approximation has to be verified first. We then computed the integrated LD profiles $g\kep$ for our different models, and used them to calculate \Vl\ and \Vtot. Results are shown in Fig.~\ref{Fig:comp} for \Vtot\ (\Vl\ are not displayed, but their behaviours are the same as \Vtot). The obtained values for the different visibilities are satisfactory and reasonably close (better than 1\%) to the full computation.

For comparison purposes, we also used the LD laws in the Kepler band published by \citet{Sing10}. We used its non-linear 4-coefficient LD laws\footnote{The improved 3-coefficient LD laws was also tested and yielded very close results. For a fair comparison, we also fitted our LD profiles with 4-coefficient laws and used them to compute visibilities: our results were almost unchanged.} to compute the visibilities. Results for \Vtot\ are also shown in Fig.~\ref{Fig:comp} (Here again, \Vl\ show exactly the same behaviour). We then see noticeable differences with the results obtained with our own LD profiles, which are greater than the difference observed between our full and approximated computations. \citet{Sing10} has also made used of the ATLAS9 code for computing LD.
The difference is mainly explained by different treatments of physical processes in the atmosphere models, especially of the convection, as already pointed out in \citet{Barban03}. We recall that \citet{Sing10} used ATLAS9 model atmospheres computed for mixing-length theory with a mixing-length parameter $\alpha=1.25$ and overshooting included, while in the present work we used the CGM approach for the convection treatment without overshooting \citep[see][for argumentation]{Heiter02}.
It shows the sensitivity of LD computations to the models and the consequences for mode-visibility calculations.

\section{Discussion}\label{sec:dis}

The notations used in this note differ from those in M09. The term $c_{\mathrm{K-bol}}$ of the present work is equal to $4/R_g$ in M09, and \Vtot\ to $R_\mathrm{osc}/R_g$. It is worth noticing that $\Vtot^2$ is also equal to the $c$ factor of \citet{Kjeldsen08}. Our results could appear to be more dependent on the metallicity than the computations of M09. This is simply an impression owing to the scales and the considered variables. For example, around $\te=6000$~K, $\Vtot^2\approx3.1$ and the spread of the different curves is around 0.06, then the corresponding dispersion for $R_\mathrm{osc}$ is around 0.07, which is consistent with the point spread observed in Fig.~7 of M09.

The coefficients introduced here appear quite naturally with the classical analysis techniques:
\begin{enumerate}
 \item Using the factor $c_{\mathrm{K-bol}}$ is straightforward: by multiplying the amplitude (for example in ppm) of an $\ell=0$ mode observed in the Kepler band, one recovers its bolometric amplitude.

\item The visibility $\Vl^2$ can be used to fix a priori ratios of mode heights in power-spectrum fitting, or to be a posteriori compared to the fitted ratios if they are free parameters.

\item In some cases, specially for stars with low signal-to-noise ratios \citep[e.g.][]{Mathur10}, the power density spectrum is convolved by a box car with a width equal to the large separation $\Delta$. The smoothed spectrum is then used to recover the power maximum $P_m$.
Assuming that one mode of all degrees is present in the interval $\Delta$ and they all have a similar intrinsic amplitude (hypothesis of energy equipartition), we recover the maximum amplitude of radial modes in the Kepler band as $A_{\ell=0,\mathrm{K}}=\sqrt{P_m/\Delta}/\Vtot$, and, then, the bolometric maximal radial-mode amplitude is $A_{\ell=0,\mathrm{bol}}=c_{\mathrm{K-bol}}\sqrt{P_m/\Delta}/\Vtot$.

\end{enumerate}

At this point, it is important to stress the limits of the kind of approach we used. As already mentioned, our bolometric corrections ignore the departure of stellar surface emissions from black body spectra. As a consequence, the effects of photospheric absorption lines and bands are neglected, as are the effects of interstellar reddening. By using a classical reddening law \citep[e.g][]{Schild77}, the bolometric correction $c_{\mathrm{K-bol}}$ is increased by less than 2\% for a reddening $E(B-V)=0.14$ --typical for the open cluster NGC6811-- and is negligible for nearby stars.
 
Second, the LD laws are only models, which are also, in essence, approximations.
It is nowadays possible to test modelled LD, at least linear approximations of LD, with various kinds of observations, such as
interferometry \citep[e.g.][]{Burns97}, light curves of eclipsing binaries \citep[e.g.][]{Popper84}, planetary transits \citep[e.g.][]{Barge08}, or microlensing \citep[e.g.][]{Fouque10}.
Besides, \citet{Sing10} has shown discrepancies between its modelled LD and the LD deduced from planetary transits observed by CoRoT. Nevertheless, new 3-D atmosphere modelling should provide better stellar LD prediction \citep[e.g.][]{Bigot06,Chiavassa09}.

Last, even considering the LD profiles are perfectly known, some other assumptions could be not fulfilled. In our approach, the physics of modes inside the atmosphere is simplified: particularly, non-adiabatic effects are neglected, and the stellar photosphere is assumed to be thin enough to render the variations of mode amplitudes with the altitude in the atmosphere negligible.
For the Sun, whose surface is resolved, we have a direct measurement of the LD at different wavelengths. It is then possible to compute visibilities \Vl\ from the real LD and to compare them to the ratios of observed mode amplitudes: \citet{Salabert11} have shown differences between the computed visibilities \Vl\ and the observations of the helioseismic VIRGO (Variability of Solar IRradiance and Gravity Oscillation) instrument. The lack of the previously mentioned physical effects in the modelled visibilities can be made responsible to explain the discrepancies. However, it is also possible that the intrinsic amplitudes of modes are different. Other stars have shown unexpected high apparent amplitudes of $\ell=3$ modes \citep[e.g.][]{Deheuvels10}. It is still unclear whether this is a consequence of incorrect computations of mode visibilities, or a signature of the over-excitation of some modes. These two examples show us how important it is to improve in the near future the computation of mode visibilities, by considering more realistic physics of acoustic modes in the stellar atmospheres, to be able to recover intrinsic mode amplitudes.

\begin{acknowledgements}
We acknowledge E. Michel for useful comments, as well as the International Space Science Institute (ISSI) for supporting the asteroFLAG international team%
\footnote{\url{http://www.issi.unibe.ch/teams/Astflag/}}, where this work was
started. We are very grateful to R.~Kurucz for making available and open to the scientific community his ATLAS9 code.
This research made use of the VizieR catalogue access tool, operated at the CDS, Strasbourg, France.
\end{acknowledgements}

\bibliographystyle{aa}
\bibliography{ref}

\end{document}